\documentclass{PoS}
\newcommand{\ba}{\begin{eqnarray}}
\newcommand{\ea}{\end{eqnarray}}
\newcommand{\rmi}[1]{{\mbox{\scriptsize #1}}}

\newcommand{\tr}{{\rm Tr\,}}
\newcommand{\nn}{\nonumber \\}

\newcommand{\msbar}{{\overline{\mbox{\rm MS}}}}

\renewcommand{\vec}[1]{{\bf #1}}

\renewcommand{\(}{\left(}
\renewcommand{\)}{\right)}

\def\Z{\mathcal{Z}}

\def\openone{\rlap 1\kern 0.22ex 1}
\newcommand{\ev}[1]{\langle #1 \rangle}
\newcommand{\heavy}{\Sigma}
\newcommand{\light}{\Pi}

\title{Center-symmetric dimensional reduction of hot Yang-Mills theory }

\ShortTitle{Center-symmetric dimensional reduction of hot Yang-Mills theory }

\author{
\speaker{Aleksi Kurkela}%
        \\
        Institut f\"ur Theoretische Physik, ETH Z\"urich, CH-8093 Z\"urich, Switzerland\\
        E-mail: \email{kurkela@phys.ethz.ch}}

\abstract{It is expected that incorporating the center symmetry in the conventional dimensionally reduced effective theory for high-temperature SU($N_c$) Yang-Mills theory, EQCD, will considerably extend its applicability towards the deconfinement transition. The construction of such a center-symmetric effective theory for the case of two colors is reviewed and lattice simulation results are presented.
The simulations demonstrate that unlike EQCD, the new center-symmetric theory undergoes a second order confining phase transition in complete analogy with the full theory.}

\FullConference{The XXVI International Symposium on Lattice Field Theory \\
		 July 14 - 19, 2008\\
		 Williamsburg, Virginia, USA}

\begin{document}

\section{Introduction}

In the study of high temperature gauge theory in thermal
equilibrium, a particularly useful approach has turned out to be
that of dimensional reduction \cite{GAP}. There, one describes the system via a
$d-1$ dimensional static effective theory built using the fact
that, at high temperature, the non-static field
modes decouple quite efficiently from the dynamics of length scales
larger than or equal to the inverse Debye mass.

How efficiently the non-static modes decouple from the dynamics of the long wavelength modes is dictated by the magnitude of the scale separation between the Debye mass and the first non-static modes. At extremely high temperatures, where the gauge coupling constant $g$ is small, the separation of scales is guaranteed as the scale associated with the non-static modes is $\sim 2\pi T$, whereas the Debye mass $\sim gT$ is suppressed by the gauge coupling. However, it is known that there exists at least a modest separation of these scales even in the vicinity of the deconfinement transition \cite{Kaczmarek}, and thus the dimensionally reduced theory should give at least a qualitative description of the full theory all the way down to $T_c$ or even below.

The perturbatively constructed dimensionally reduced effective theory Electrostatic QCD (or EQCD), an effective theory for the zero Matsubara modes of gauge fields $A_\mu$, has had many successes in the high temperature regime, such as the efficient reorganization of the weak coupling expansion of the QCD pressure \cite{KLRS}. This reorganization has provided a framework for extending the expansion to the full $g^6$ order, where the pressure acquires its first non-perturbative contributions. In addition to this, there are several numerical simulation results from EQCD, which have produced results matching those of the full four-dimensional theory even at surprisingly low temperatures all the way down to $\sim 1.5T_c$ \cite{HLP}.\footnote{The spatial string tension of (in 2+1 flavor QCD) seems to be described very well by EQCD even further, to temperatures very near $T_c$ \cite{Laermann}.} 

However, even with these successes, EQCD cannot accommodate the approach to $T_c$ as in EQCD the dynamics responsible for the phase transition are missing. Being a perturbatively constructed effective theory, EQCD describes small fluctuations around one of the $N_c$ (in the quarkless case degenerate) deconfining minima, whereas the qualitative change near $T_c$ is closely related with the tunnelings of the Polyakov loop between the different deconfining phases. Consequences of this shortcoming are seen for example in the phase diagram of the effective theory: The phase in EQCD corresponding to the physical deconfined phase is not the global minimum of the effective theory and simulations have to be performed in a metastable phase, discarding by hand the contributions of the global minima of the theory to the partition function \cite{KLRS2}. 

In order for the effective theory to correctly describe the dynamics of the large field fluctuations, it has to accommodate the full symmetry structure of the underlying theory, which in this case includes the ${\rm Z}_N$ center symmetry of the Yang-Mills theory\footnote{Even though in the full QCD the quarks break the center symmetry softly preferring the real deconfined minimum, the metastable minima contribute to the partition function and should be accounted for near the deconfinement transition.}. A natural way to construct an effective theory with the center symmetry is to use some remnant of the temporal Wilson line as a degree of freedom instead of the small fluctuations of the temporal gauge field around a deconfining minimum.
Such a center-symmetric effective theory for SU(3) Yang-Mills theory was proposed in \cite{VY} and further formulated on a lattice in \cite{AK}. Subsequently, in order to create a more economical platform to study the role of the center symmetry, a center-symmetric effective theory for SU(2) Yang-Mills was constructed and studied in \cite{FKV}. The restriction for the degree of freedom in the theory to lie on the SU($N_c$) manifold makes it impossible to construct a super-renormalizable theory with polynomial interactions\footnote{A non-renormalizable effective theory of temporal Wilson lines is studied in \cite{P}.}, and thus these theories are formulated using the spatially \emph{coarse grained} temporal Wilson line as the degree of freedom.

The main result from the simulation of the SU(2) case is that upon the inclusion of the center symmetry, the effective theory accommodates a confining second-order phase transition in the same (3d-Ising) universality class as the full four-dimensional theory, which happens at an effective theory coupling consistent with the critical coupling of the full theory.

\section{Center-symmetric Lagrangian}
The center-symmetric effective theory for hot SU($N_c$) Yang-Mills theory is defined by the action
\begin{eqnarray}
    S&=&\int d^3 x \mathcal{L}(x), \\
    \mathcal{L}
    &=&
    g_3^{-2}\Big\{\frac{1}{2} \, \tr  F_{ij}^2
    + \tr\! \left(D_i \Z^{\dagger}D_i\Z\right)
    + V(\Z)\Big\}
    ,
\label{lageff2}
\end{eqnarray}
with $D_i \equiv \partial_i -i [A_i,\,\cdot\,]$
and $F_{ij} \equiv \partial_i A_j - \partial_j A_i - [A_i,A_j]$, $i,j=1,2,3$. To leading order, the fields $A_i$ are the zero Matsubara modes of the four-dimensional theory whereas the field $\Z$ is the (gauge invariantly) coarse grained temporal Wilson line
\ba
\Z(\vec{x})=\frac{T}{V_\rmi{Block}}\int_V d^3y \,
U(\vec{x},\vec{y})\Omega(\vec{y})U(\vec{y},\vec{x}).
\ea
Here the integration goes over the (somewhat arbitrary) ${\mathcal O}(T^{-3})$ volume of the block and
$U(\vec{x},\vec{y})$ is a parallel transport connecting the points $\vec{x}$
and $\vec{y}$ at constant time $\tau=0$, whereas $\Omega(\vec{x})$ is the ordinary temporal Wilson line winding around the Euclidean time direction
\begin{eqnarray}
    \Omega(\mathbf{x})
    &\equiv&
     {\cal P} \, \exp
    \bigg[i \int_0^{\beta}\!\!{\rm d\tau} \> A_0(\tau,\mathbf{x})\bigg].
\label{omega}
\end{eqnarray}

In the case of the SU(2), the $2\times2$ coarse grained temporal Wilson line can be expressed by using the scalar fields $\Sigma$ and $\Pi_a$ ($a=1,2,3$) and Pauli matrices $\sigma_a$
\ba
    \Z = \frac{1}{2}\Big\{\heavy \openone + i \light_a \sigma_a \Big\},
\ea
and the potential $V(\Z)$, consisting of all other possible super-renormalizable operators constructed from the fields in the effective theory respecting the symmetries of the full theory, can be expressed as
\begin{eqnarray}
V(\Z)&=&b_1 \heavy^2 + b_2 \light_a^2 +c_1 \heavy^4 + c_2 \(\light_a^2\)^2 + c_3 \heavy^2 \light_a^2 \label{pot1}.
\end{eqnarray}

The parameters of the effective theory are related to those of the full theory by imposing conditions that at leading order the effective theory reduces to EQCD at high temperatures and that a domain wall stretching from one deconfined minimum to another has the correct tension, resulting in 
\ba
b_1&=&-\frac{1}{4}r^2 T^2,  \\ b_2&=&-\frac{1}{4}r^2T^2+0.441841 g^2T^2,\\
c_1&=&0.0311994r^2+0.0135415g^2,\\
c_2&=&0.0311994 r^2+0.008443432g^2,\\
c_3&=&0.0623987r^2, \label{c3num}\\
g_3^2 &=& g^2 T \label{gmatch},
\ea
where $g$ and $T$ are the coupling constant and temperature of the four-dimensional theory, and $rT$ is an $\mathcal{O}(T)$ mass scale associated with the auxiliary scalar field introduced by the coarse graining. This quantity, closely related to the cutoff of the effective theory, is not perturbatively matched and the dynamics of the long wavelength modes should not be affected by its specific value. 

\section{Non-perturbative phase diagram of the effective theory}
Since the effective theory is super-renormalizable, the theory can be formulated on a lattice and the lattice-continuum relations of the parameters of the Lagrangians can be computed up to and including $\mathcal{O}(a^0)$ using two-loop lattice perturbation theory, making it possible to simulate the theory on a lattice at the physical $\msbar$ parameters \cite{AK,FKV}. Using standard Wilson discretization and denoting lattice quantities with hats, the lattice action corresponding to effective theory reads
\ba
S_a &=& S_W + S_\Z + V(\hat{\heavy},\hat{\light}), \label{XX}\\
S_W &=&  \beta \sum_{x,i<j} \left[ 1 - \frac{1}{2} \tr[ U_{ij}] \right],\\
S_\Z &=&  2 \(\frac{4}{\beta}\) \sum_{x,i}\tr \left[\hat{\light}^2 -
\hat{\light}(x)U_i(x)\hat{\light}(x+\hat{i})U^\dagger_i(x)\right] \nn
&&\;\;\;+  \(\frac{4}{\beta}\)\sum_{x,i}\left(\hat{\heavy}^2(x)- \hat{\heavy}(x)\hat{\heavy}(x+\hat{i})\right),\\
V &=& \(\frac{4}{\beta}\)^3 \sum_{x} \left[ \hat{b}_1 \hat{\heavy}^2 + \hat{b}_2 \hat{\light}_a^2
+ \hat{c}_1 \hat{\heavy}^4 + \hat{c}_2 \(\hat{\light}_a^2\)^2 + \hat{c}_3 \hat{\heavy}^2 \hat{\light}_a^2 \right],
\ea
where $\beta$ is the lattice coupling constant
\ba
\beta &=& \frac{4}{a g_3^2}
\ea
corresponding to a lattice spacing $a$. The lattice quantities are related to the continuum $\msbar$ quantities via
\begin{eqnarray}
\hat{\heavy} = \heavy/g_3+ \mathcal{O}(\beta^{-1}),\,\,\,\,\,
\hat{\light} = \light/g_3+ \mathcal{O}(\beta^{-1}), \,\,\,\,\,
\hat{c}_i &= c_i+ \mathcal{O}(\beta^{-1}),
\label{eq:tree}
\end{eqnarray}
and
\ba
  \hat{b}_1 &= &\;b_1/g_3^4-\frac{2.38193365}{4\pi}(2c_1  + c_3)\beta \nn
&&+  \frac{1}{16\pi^2}\left\{(48c_1^2+12c_3^2-12c_3)\left[\log1.5\beta + 0.08849 \right] -6.9537\,c_3\right\}+\mathcal{O}(\beta^{-1})),\label{eq:2loop_b1n}\\
\hat{b}_2& = & \;b_2/g_3^4-\frac{0.7939779}{4\pi}(10c_2+c_3+2)\beta\label{eq:2loop_b2n}\\
&&+\frac{1}{16\pi^2}\bigg\{(80c_2^2 + 4c_3^2-40c_2)\left[\log1.5\beta+0.08849 \right]-23.17895\,c_2
-8.66687\bigg\}+ \mathcal{O}(\beta^{-1}).\nonumber
\ea

\begin{figure}[ht]
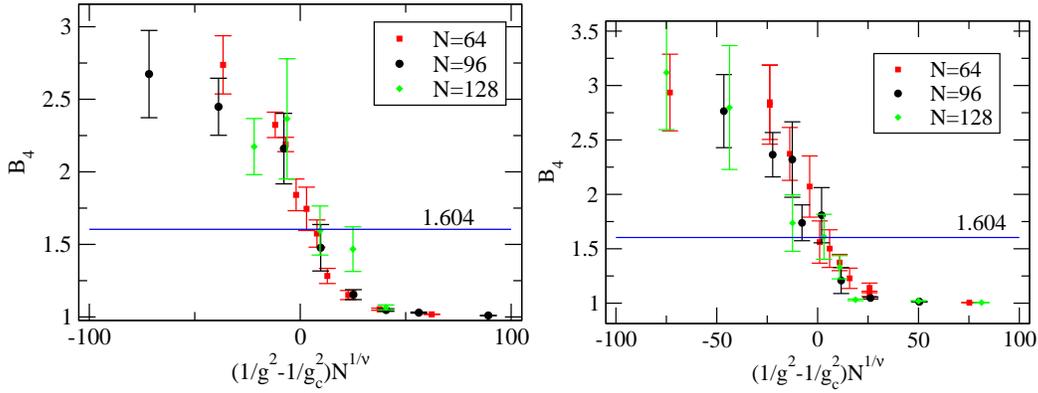

\begin{center}
\includegraphics*[width = 0.46\textwidth]{binder4r5.eps}
\includegraphics*[width = 0.44\textwidth]{binder4r10.eps}
\caption{A check of the universality class of the transition:
The Binder cumulant $\langle(\tr\Z)^4\rangle / \langle(\tr\Z)^2\rangle^2$
is shown for various lattice volumes $N^3$ as a function of the rescaled variable $(1/g^2 - 1/g^2_c) N^{1/\nu}$
for $r^2=5$ (left) and $10$ (right).
The pseudocritical coupling $g^2_c$ is the value of the coupling, which maximizes the susceptibility of $\tr\Z$, and $\nu=0.63$ as appropriate for a three-dimensional Z(2)-transition.
A satisfactory data collapse is observed for various volumes, and the cumulant value
at $g^2_c$ is consistent with the 3d-Ising value 1.604.}
\label{fig:B4}\end{center}
\end{figure}
 
The non-perturbative phase diagram at any fixed $r$ closely resembles that of the four-dimensional SU(2) Yang-Mills theory: There are three phases, the two deconfined phases with $\ev{\tr \Z}\neq0$ which occur at small $g$, and the remnant of the confined phase with $\ev{\tr \Z}=0$, seen at large $g$. 
The confined and deconfined phases are separated by a second order transition, which belongs to the universality class of 3d-Ising model (see Fig.~1), the correct universality class of the four-dimensional theory. 

\begin{figure}[ht]
\begin{center}
\includegraphics*[width = 0.5\textwidth]{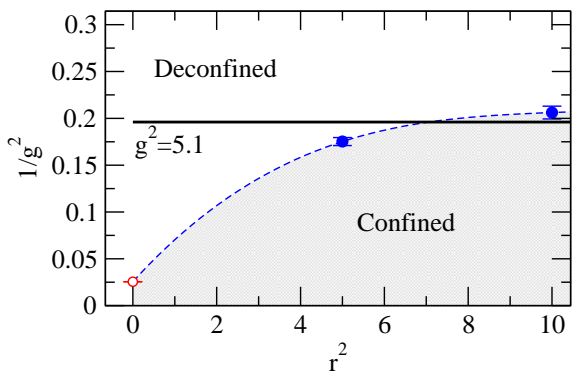}
\caption{The phase diagram of the effective theory on the $(r^2,1/g^2)$ plane. The solid blue data points are from numerical simulations from \cite{FKV}, and the open red one has been obtained from the known location of the critical point of $\lambda\phi^4$ theory. The curve connecting the points has been obtained from a polynomial fit and has been included in the diagram to guide the eye. The horizontal line is the critical coupling of the full four-dimensional theory.} \label{fig:phasediagram} \end{center}
\end{figure}

The phase diagram in the $(r,g)$ plane is depicted in Fig.~2. As $rT$ becomes of order $T$, the phase diagram depends only mildly on $r$, as expected. Remarkably, at large values of $r$, the critical coupling of the effective theory is even consistent with the four-dimensional theory critical coupling (using one-loop running to convert the critical temperature in lattice units to $\msbar$ coupling).

\section{Conclusions}
It has been seen that in the case of SU(2) Yang-Mills theory, the accommodation of the ${\rm Z}_N$ center symmetry in the dimensionally reduced effective theory improves the applicability of the effective theory near the deconfinement transition significantly. Lattice simulations show, that upon respecting the symmetries of the full theory, the phase diagram of the effective theory becomes qualitatively similar to that of the four-dimensional theory having two deconfined phases at high temperature and a confined phase at low temperature separated by a second order transition in 3d-Ising universality class. In addition to this, quantitatively the phase transition takes place at effective coupling consistent with the critical coupling of the four-dimensional theory.

The success of implementing the center symmetry to the SU(2) case encourages further studies. The accuracy of the effective theory near the deconfining transition should be quantified for example by studying the behavior of various screening masses and by measuring the non-perturbative domain wall profile. In addition to this, the effective theory can be extended to give predictions of physical situations which are otherwise difficult to study. These include for example the possibility to study the poorly known region of the phase diagram of QCD with heavy quarks by including center symmetry breaking operators to the Lagrangian, and extensions to large $N$, a work which is already started in \cite{Korthals}.

\end{document}